\documentclass{elsart}
\usepackage[dvips]{graphicx}
\usepackage{eurosym}
\begin{document}


\begin{frontmatter}
\title{The Power-law Tail Exponent of Income Distributions}
\author[Roma,Canberra]{F. Clementi\corauthref{cor}},
\corauth[cor]{Corresponding author: Tel.: +39--06--49766843; fax: +39--06--4461964.}
\ead{fabio.clementi@uniroma1.it}
\author[Canberra]{T. Di Matteo},
\author[Ancona]{M. Gallegati}
\address[Roma]{Department of Public Economics, University of Rome ``La Sapienza'', Via del Castro Laurenziano 9, 00161 Rome, Italy}
\address[Canberra]{Applied Mathematics, Research School of Physical Sciences and Engineering, The Australian National University, 0200 Canberra, Australia}
\address[Ancona]{Department of Economics, Universit\`a Politecnica delle Marche, Piazzale Martelli 8, 60121 Ancona, Italy}
\begin{abstract}
In this paper we tackle the problem of estimating the power-law tail exponent of income distributions by using the Hill's estimator. A subsample semi-parametric bootstrap procedure minimising the mean squared error is used to choose the power-law cutoff value optimally. This technique is applied to personal income data for Australia and Italy.
\end{abstract}
\end{frontmatter}


\section{Introduction}
Since Pareto it has been recognized that a \textit{power-law} provides a good fit for the distribution of high incomes \cite{Pareto1897}. The Pareto's law asserts that the complementary cumulative distribution $P_{>}\left(y\right)=1-\int_{-\infty}^{y}p\left(\xi\right)\,\mathrm{d}\xi\rightarrow P_{>}\left(u\right)\left(\frac{u}{y}\right)^{\alpha}$, with $y\geq u$, where $u>0$ is the threshold value of the distribution and $\alpha>0$ turns out to be some kind of index of inequality of distribution. The fit of such distribution is usually performed by judging the degree of linearity in a double logarithmic plot involving the empirical and theoretical distribution functions, in such a way that the estimation of $u$ of the distribution does not seem to follow a neutral procedure. Moreover, recent studies have criticized the reliability of this geometrical method by showing that linear-fit based methods for estimating the power-law exponent tend to provide biased estimates, while the maximum likelihood estimation method produces more accurate and robust estimates \cite{GoldsteinMorrisYen2004,BrizioMontoya2005}. Hill proposed a conditional maximum likelihood estimator for $\alpha$ based on the $k$ largest order statistics for non-negative data with a Pareto's tail \cite{Hill1975}. That is, if $y_{\left[n\right]}\geq y_{\left[n-1\right]}\geq\ldots\geq y_{\left[n-k\right]}\geq\ldots\geq y_{\left[1\right]}$, with $y_{\left[i\right]}$ denoting the $i$\textsuperscript{th} order statistic, are the sample elements put in descending order, then the Hill's estimator is
\begin{equation}
\hat{\alpha}_{n}\left(k\right)=\left[\frac{1}{k}\sum\limits_{i=1}^{k}\left(\log y_{n-i+1}-\log y_{n-k}\right)\right]^{-1}
\label{eq:HE}
\end{equation}
where $n$ is the sample size and $k$ an integer value in $\left[1,n\right]$. Unfortunately, the finite-sample properties of the estimator (Eq.~\ref{eq:HE}) depend crucially on the choice of $k$: increasing $k$ reduces the variance because more data are used, but it increases the bias because the power-law is assumed to hold only in the extreme tail.
\par
Over the last twenty years, estimation of the Pareto's index has received considerable attention in extreme value statistics \cite{Lux2001}. All of the proposed estimators, including the Hill's estimator, are based on the assumption that the number of observations in the upper tail to be included, $k$, is known. In practice, $k$ is unknown; therefore, the first task is to identify which values are really extreme values. Tools from exploratory data analysis, as the quantile-quantile plot and/or the mean excess plot, might prove helpful in detecting graphically the quantile $y_{\left[n-k\right]}$ above which the Pareto's relationship is valid; however, they do not propose any formal computable method and, imposing an arbitrary threshold, they only give very rough estimates of the range of extreme values.
\par
Given the bias-variance \textit{trade-off} for the Hill's estimator, a general and formal approach in determining the best $k$ value is the minimisation of the \textit{Mean Squared Error} (\textit{MSE}) between $\hat{\alpha}_{n}\left(k\right)$ and the theoretical value $\alpha$. Unfortunately, in empirical studies of data the theoretical value of $\alpha$ is not known. Therefore, an attempt to find an approximation to the sampling distribution of the Hill's estimator is required. To this end, a number of innovative techniques in the statistical analysis of extreme values proposes to adopt the powerful bootstrap tool to find the optimal number of order statistics adaptively \cite{Hall1990,DacorognaMullerPictetDeVries1992,DanielssonDeHaanPengDeVries2001,Lux2000}. By capitalizing on these recent advances in the extreme value statistics literature, in this paper we adopt a subsample semi-parametric bootstrap algorithm in order to make a reasonable and more automated selection of the extreme quantiles useful for studying the upper tail of income distributions and to end up at less ambiguous estimates of $\alpha$. This methodology is described in Section~\ref{sec:EstimationTechniqueForThresholdSelection} and its application to Australian and Italian income data ~\cite{DiMatteoAsteHyde2004,ClementiGallegati2005} is given in Section~\ref{sec:EmpiricalApplicationTheAustralianAndItalianPersonalIncomeDistributions}. Some conclusive remarks are reported in Section~\ref{sec:ConcludingRemarks}.


\section{Estimation Technique for Threshold Selection}
\label{sec:EstimationTechniqueForThresholdSelection}
In this section we consider the problem of finding the optimal threshold $u_{n}^{\ast}$ -- or equivalently the optimal number $k^{\ast}$ of extreme sample values above that threshold -- to be used for estimation of $\alpha$. In order to achieve this task, we minimize the \textit{MSE} of the Hill's estimator (Eq.~\ref{eq:HE}) for a series of thresholds $u_{n}=y_{\left[n-k\right]}$, and pick the $u_{n}$ value at which the \textit{MSE} attains its minimum as $u_{n}^{\ast}$. Given that different threshold series choices define different sets of possible observations to be included in the upper tail of a specific observed sample $\mathbf{y}_{n}=\left\{y_{i};i=1,2,\ldots,n\right\}$, only the observations exceeding a certain threshold that are additionally distributed according to a Pareto's cumulative distribution function $PD_{\hat{\alpha}_{n}\left(k\right),u_{n}}\left(y\right)$ are included in the series. In order to check this condition, we perform for each threshold in the original sample a \textit{Kolmogorov-Smirnov} (\textit{K-S}) goodness-of-fit test for the null hypothesis $H_{0}:\hat{F}_{n}\left(y\right)=PD_{\hat{\alpha}_{n}\left(k\right),u_{n}}\left(y\right)$ versus the general alternative of the form $H_{1}:\hat{F}_{n}\left(y\right)\neq PD_{\hat{\alpha}_{n}\left(k\right),u_{n}}\left(y\right)$,
where $\hat{F}_{n}\left(y\right)$ is the empirical distribution function, and $\hat{\alpha}_{n}\left(k\right)$ is a prior estimate for each threshold $u_{n}$ of the Pareto's tail index obtained through the Hill's statistic. Following the methodology in \cite{Stephens1974}, the formal steps in making a test of $H_{0}$ are as follows:
\renewcommand{\theenumi}{\alph{enumi}}
\begin{enumerate}
\item Calculate the original \textit{K-S} test statistic $D$ by using the formula $D=\sup\limits_{-\infty<y<\infty}\left|\hat{F}_{n}\left(y\right)-PD_{\hat{\alpha}_{n}\left(k\right),u_{n}}\left(y\right)\right|$.
\item Calculate the modified form $T^{\ast}$ by using the formula
\begin{equation}
T^{\ast}=D\left(\sqrt{n}+0.12+\frac{0.11}{\sqrt{n}}\right).
\label{eq:KS}
\end{equation}
\item Reject $H_{0}$ if $T^{\ast}$ exceeds the cutoff level, $z$, for the chosen significance level.
\end{enumerate}
To obtain an estimate of finite-sample bias and variance (and thus \textit{MSE}) at each threshold coming from the null hypothesis $H_{0}$, a natural criterion is to use the \textit{bootstrap} \cite{Efron1979}. In its purest form, the bootstrap involves approximating an unknown distribution function, $F\left(y\right)$, by the empirical distribution function, $\hat{F}_{n}\left(y\right)$. However, most times the empirical distribution model from which one resamples in a purely non-parametric bootstrap is not a good approximation of the distribution shape in the tail. Therefore, we initially smooth the tail data by fitting a Pareto's cumulative distribution function
\begin{equation}
PD_{\hat{\alpha}_{n}\left(k\right),u_{n}}\left(y\right)=p=1-P_{>}\left(u_{n}\right)\left(\frac{u_{n}}{y}\right)^{\hat{\alpha}_{n}\left(k\right)}
\label{eq:PDF}
\end{equation}
to the $n_{1}\leq n$ observations $\mathbf{y}_{n_{1}}=\left\{y\in\mathbf{y}_{n}:T^{\ast}\leq z\right\}$, and then use the quantiles $\mathbf{y}^{p}_{n_{1}}=\left\{y\in\mathbf{y}_{n_{1}}:PD_{\hat{\alpha}_{n}\left(k\right),u_{n}}\left(y\right)\geq p\right\}$ obtained directly from inverting the estimated model (Eq.~\ref{eq:PDF}) to draw the bootstrap samples.
\par
Let us here summarize the adopted methodology:
\renewcommand{\theenumi}{\arabic{enumi}}
\begin{enumerate}
\item Evaluate the estimate $\hat{\alpha}_{n}\left(k\right)$ of the Pareto's tail index for each threshold in the original sample $\mathbf{y}_{n}$ by using the Hill's estimator (Eq.~\ref{eq:HE}).
\item For each threshold in the original sample, test the Pareto's approximation by computing the value of the \textit{K-S} test statistic (Eq.~\ref{eq:KS}).
\item Fit the model (Eq.~\ref{eq:PDF}) to the subset of data $\mathbf{y}_{n_{1}}$ belonging to the null hypothesis $H_{0}$.
\item Select $R$ independent bootstrap samples $\mathbf{y}^{\#}_{1},\mathbf{y}^{\#}_{2},\ldots,\mathbf{y}^{\#}_{R}$, each consisting of $n_{1}$ values drawn with replacement from the set of quantiles $\mathbf{y}^{p}_{n_{1}}$ obtained by inverting the fitted model (Eq.~\ref{eq:PDF}).
\item For each bootstrap sample $\mathbf{y}^{\#}_{r}$, $r=1,2,\ldots,R$, and for each threshold $u^{\#}_{n_{1}}$ in the bootstrap sample, evaluate the bootstrap estimate $\hat{\alpha}^{\#}_{n_{1}}\left(k_{1}\right)$ of the Pareto's tail index by using the Hill's estimator (Eq.~\ref{eq:HE}).
\item For each threshold $u^{\#}_{n_{1}}$, calculate the bias, $B=E\left[\hat{\alpha}^{\#}_{n_{1}}\left(k_{1}\right)\right]-\hat{\alpha}_{n}\left(k\right)$, the variance, $Var=E\left\{\left[\hat{\alpha}^{\#}_{n_{1}}\left(k_{1}\right)\right]^{2}\right\}-\left\{E\left[\hat{\alpha}^{\#}_{n_{1}}\left(k_{1}\right)\right]\right\}^{2}$, and the mean squared error, $MSE=B^{2}+Var$, of the Hill's tail index estimates.
\item Select as the optimal threshold $u^{\ast}_{n}=y_{\left[n-k^{\ast}\right]}$ that threshold where the \textit{MSE} attains its minimum.
\end{enumerate}
Minimising the \textit{MSE}, thus, amounts to find the \textit{MSE} minimising number of order statistics $k^{\ast}=\arg\min\limits_{k}MSE$, from which one infers the optimal estimate of the tail index $\hat{\alpha}^{\ast}_{n}\left(k^{\ast}\right)$.


\section{Empirical Application: The Australian and Italian Personal Income Distributions}
\label{sec:EmpiricalApplicationTheAustralianAndItalianPersonalIncomeDistributions}
The data sources we use to illustrate how the methodology proposed in Section \ref{sec:EstimationTechniqueForThresholdSelection} can be applied to the analysis of income distributions have been selected from the nationally representative cross-sectional data samples of the Australian and Italian household populations. In particular, we have analyzed the Total annual income from all sources in the years 1993--94 to 1996-—97, and then in 1989--90, 1998--99, 1999--2000, and 2001—-02 for Australia, and 1977--2002 for Italy~\cite{DiMatteoAsteHyde2004,ClementiGallegati2005,ClementiDiMatteoGallegati2006}. Here we report only the results in the year 1999—-2000 for Australia and 2000 for Italy.
\par
\begin{figure}
\centering
\includegraphics[width=0.49\textwidth]{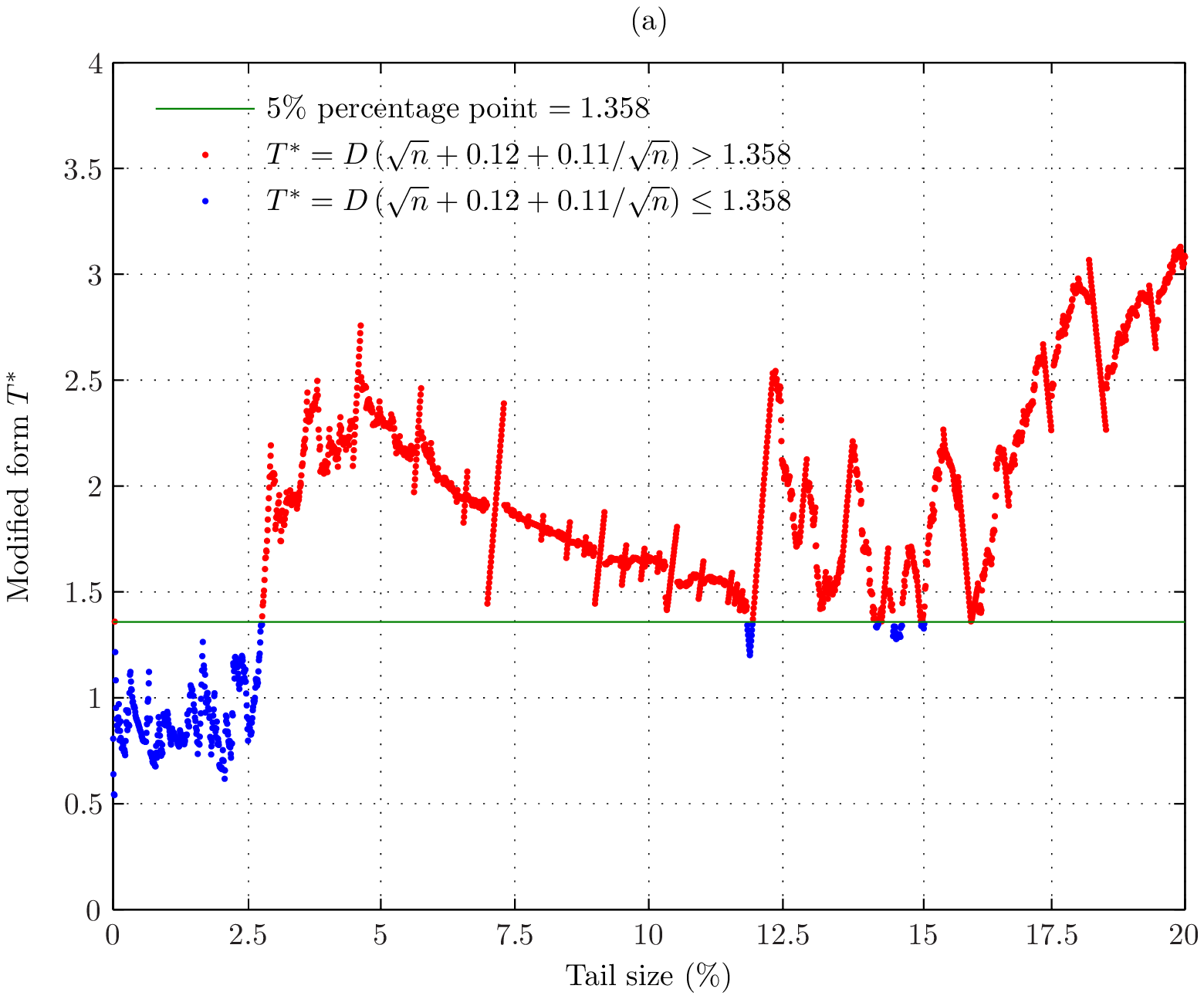}
\includegraphics[width=0.49\textwidth]{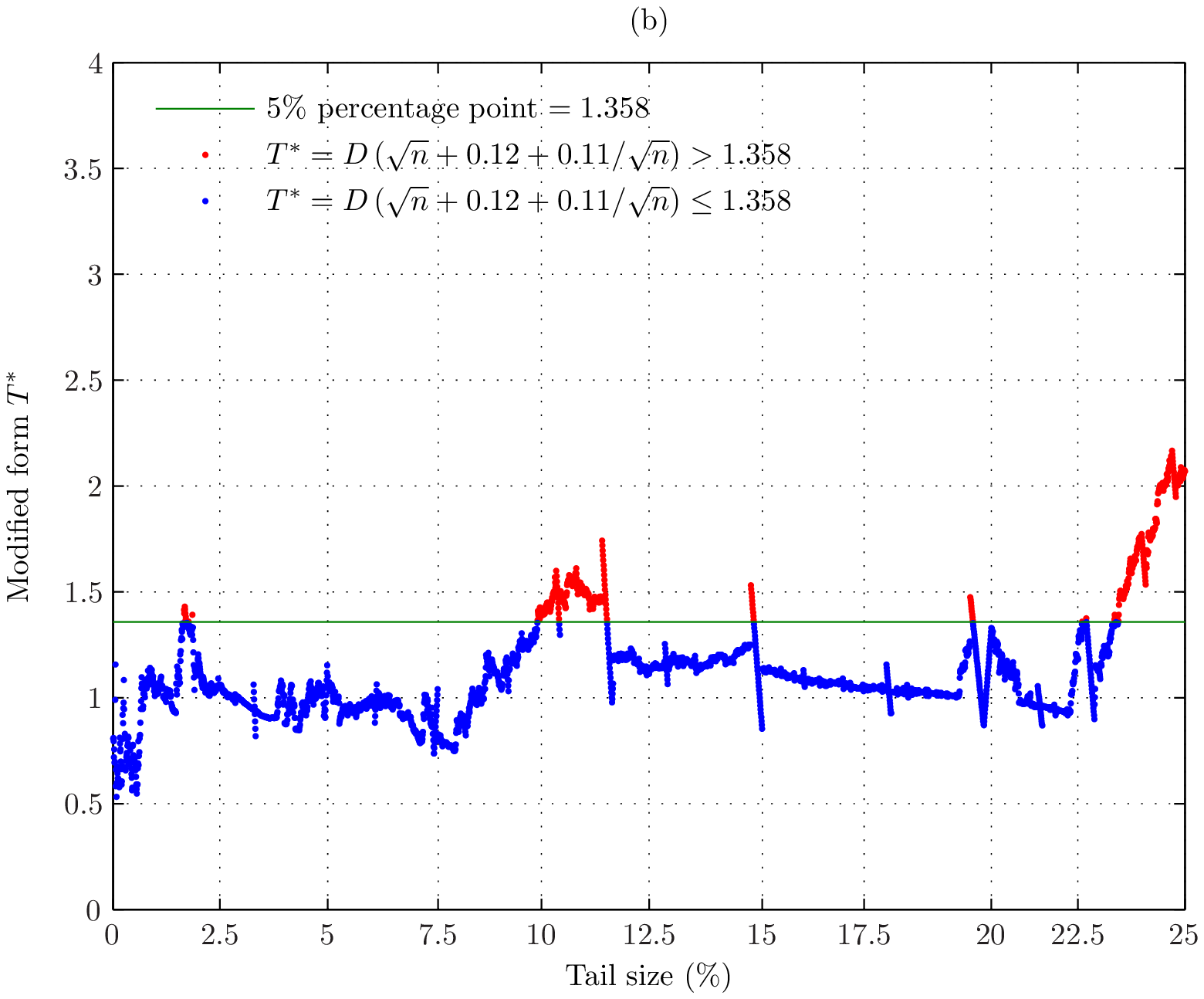}
\caption{Modified \textit{K-S} statistic (Eq.~\ref{eq:KS}) as a function of the tail size for (a) Australia in 1999--2000 and (b) Italy in 2000.}
\label{fig:Fig1}
\end{figure}

Figs.~\ref{fig:Fig1} (a) and (b) depict the outcomes of the complete sequences of \textit{K-S} test for a selection of tail fractions.
Blue points (see on line version) mark all the observations for which the modified \textit{K-S} statistic (Eq.~\ref{eq:KS}) does not exceed the $5\%$ cutoff level $z=1.358$ (solid lines in the figures). The 5\% significance point $z=1.358$ comes from Table \textit{1A} in~\cite{Stephens1974}. The figures indicate the tail regions that may be tentatively regarded as appropriate for the implementation of the semi-parametric bootstrap technique.
\par
\begin{figure}
\includegraphics[width=0.49\textwidth]{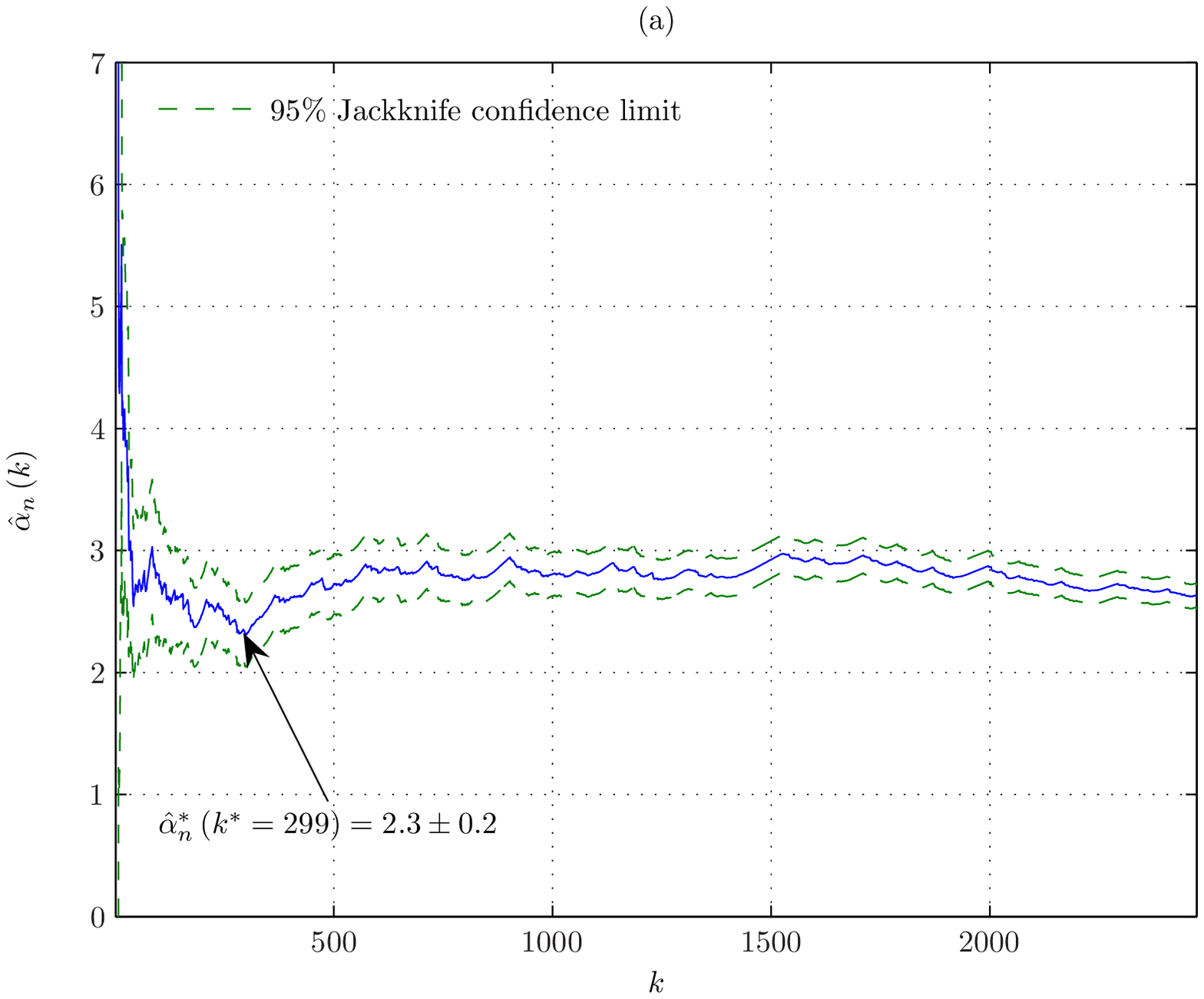}
\includegraphics[width=0.49\textwidth]{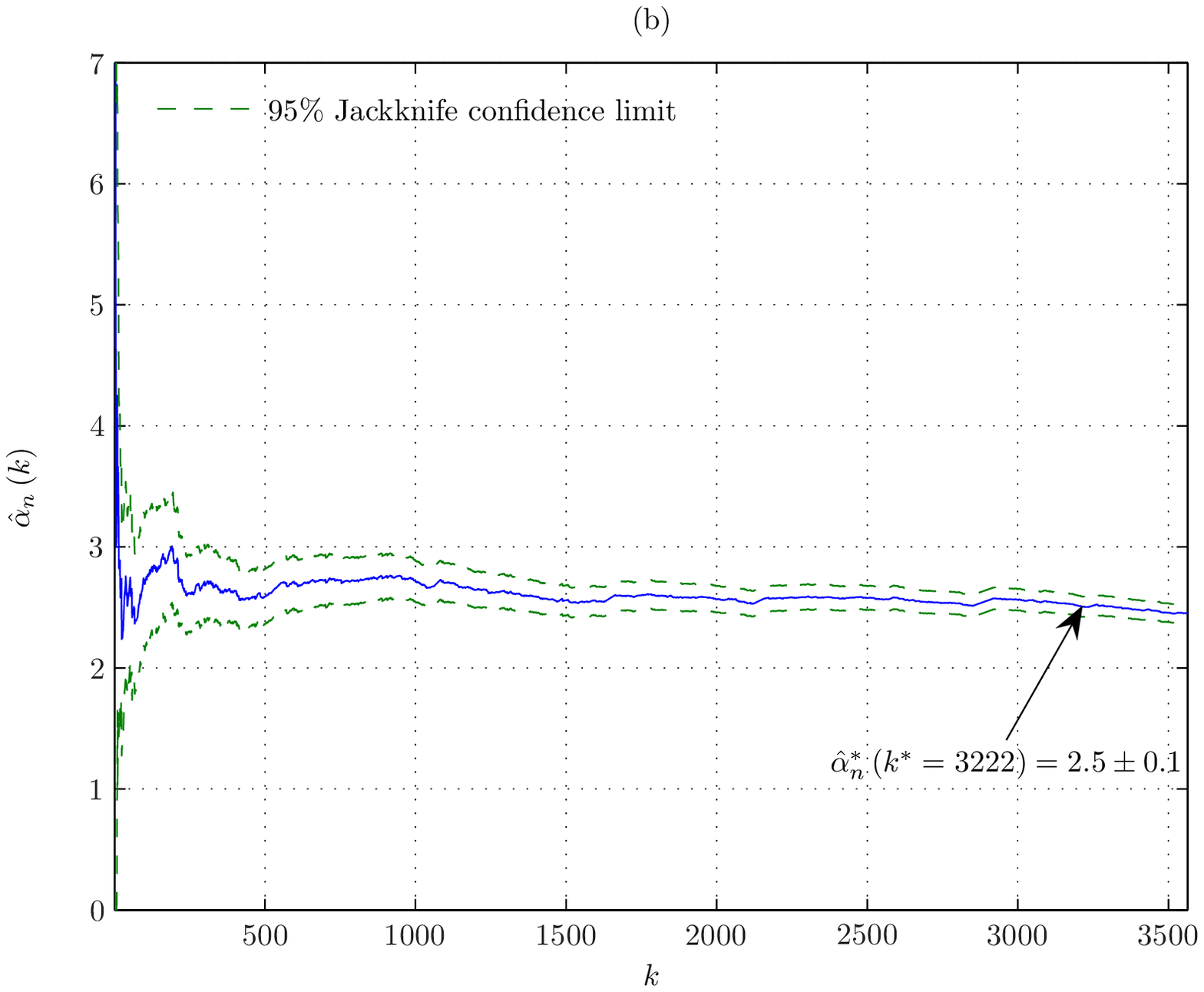}
\caption{The Hill's estimator (Eq.~\ref{eq:HE}) for (a) Australia in 1999--2000 and (b) Italy in 2000. The dashed lines represent the 95\% confidence limits of the tail index estimates computed by using the jackknife method. The arrows mark the optimal number of extreme sample values $k^{\ast}$.}
\label{fig:Fig2}
\end{figure}

The Hill's estimator (Eq.~\ref{eq:HE}) is reported in Figs.~\ref{fig:Fig2} for Australia (a) and Italy (b), and for tails $\leq20\%$ and $\leq25\%$ of the full sample size respectively (see solid lines). In these figures, the optimal number of extreme sample values are reported, namely $k^{\ast}=299$ for Australia and $k^{\ast}=3222$ for Italy, providing the following values for the tail power-law exponents: $\hat{\alpha}^{\ast}_{n}\left(k^{\ast}\right)=2.3\pm0.2$ and $\hat{\alpha}^{\ast}_{n}\left(k^{\ast}\right)=2.5\pm0.1$, where the errors (with 95\% confidence) have been obtained through the \textit{jackknife} method~\cite{PictetDacorognaMuller1996}. In these computations, we have used 1000 resamples and the subsample size has been set equal to the number of observations not rejected by the \textit{K-S} test at the 5\% level (see Section~\ref{sec:EstimationTechniqueForThresholdSelection} and Figs.~\ref{fig:Fig1} (a) and (b)). Repeated calculations with a different number of replications produce a spread of tail index estimates with deviations inside the 95\% uncertainty band (dashed lines in the figures), showing therefore numerical robustness of our results. We have here obtained more precise values of the power-law tails than the previous one reported in the literature~\cite{ClementiGallegati2005}.
\par
\begin{figure}
\includegraphics[width=0.49\textwidth]{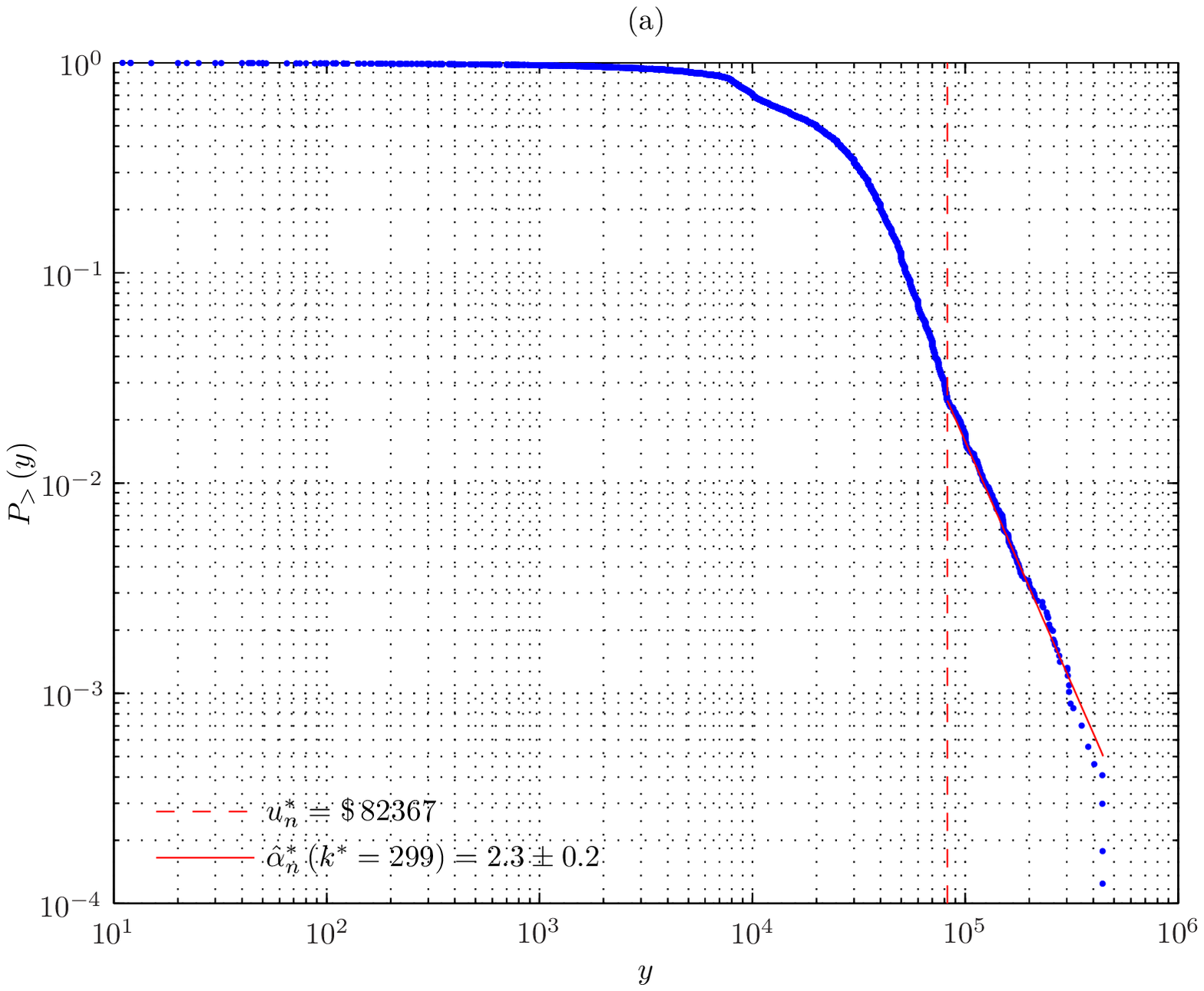}
\includegraphics[width=0.49\textwidth]{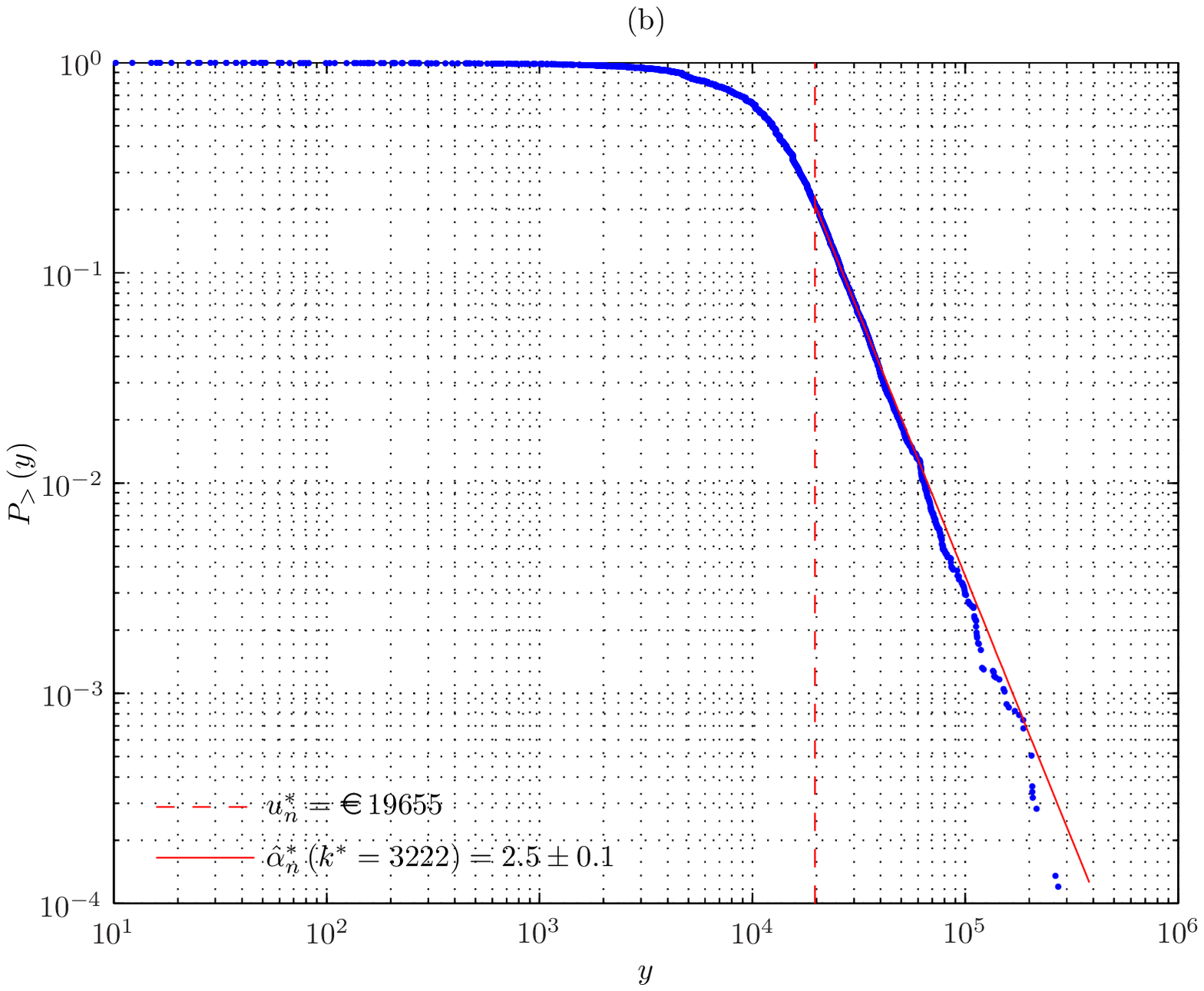}
\caption{Complementary cumulative distribution (a) for Australia in 1999-2000 and (b) for Italy in 2000 and power-law fits by using the estimated optimal values for $\alpha$.}
\label{fig:Fig3}
\end{figure}

The use of these $\hat{\alpha}^{\ast}_{n}$ optimal values produces the fits shown by the solid lines in Figs.~\ref{fig:Fig3} (a) and (b) for Australia and Italy, where the complementary cumulative distributions are plotted on a log-log scale. The vertical dashed lines indicate the optimal values of the threshold parameter attained by subsample semi-parametric bootstrapping: (a) $u^{\ast}_{n}=\$\,82367$ for Australia in 1999-2000 and (b) $u^{\ast}_{n}=\;$\EUR{19655} for Italy in 2000. As we can see, our procedure succeeds in avoiding deviations from linearity for the largest observations that might strongly influence the estimation of $\alpha$, illustrating therefore the importance of optimally choosing the tail threshold.


\section{Concluding Remarks}
\label{sec:ConcludingRemarks}
In this paper we have considered the problem of the estimation of the power-law tail exponent of income distributions and we have adopted a subsample semi-parametric bootstrap procedure in order to arrive at less ambiguous estimates of $\alpha$. This methodology has been empirically applied to the estimation of personal income distribution data for Australia and Italy. The reliability and robustness of the results have been tested by running different repeated bootstrap replications and comparing the variability of the estimates through a jackknife method.
\par
From the economic point of view, this technique for the estimation of the Pareto's tail index of income distribution is expected to allow a deeper understanding of both the way in which cyclical fluctuations in economic activity affect factor income shares and the channels through which these effects work through the size distribution of income, which are issues of relevance for the modeling of the income process in the high-end tail of the distribution.


\begin{ack}
T. Di Matteo wishes to thank the Australian Social Science Data Archive, ANU, for providing the ABS data and the partial support by ARC Discovery Projects: DP03440044 (2003) and DP0558183 (2005), COST P10 ``Physics of Risk'' project and M.I.U.R.-F.I.S.R. Project ``Ultra-high frequency dynamics of financial markets''.
\end{ack}


\end{document}